\documentclass[12pt]{iopart}

\usepackage{iopams}  

\usepackage{graphicx}
\usepackage{color}

\begin{document}

\def\tr{\text{tr}}
\def\Tr{\mbox{Tr}}

\newtheorem{theo}{Theorem}

\title{Fluctuation Relation for Quantum Heat Engines and Refrigerators}
\author{Michele Campisi}
\address{NEST, Scuola Normale Superiore \& Istituto Nanoscienze-CNR, I-56126 Pisa, Italy}
\date{\today }

\begin{abstract}
At the very foundation of the second law of thermodynamics lies the fact that no heat engine operating between two reservoires of temperatures $T_C\leq T_H$ can overperform the ideal Carnot engine: $\langle W \rangle / \langle Q_H \rangle \leq 1-T_C/T_H$. This inequality follows from an exact  fluctuation relation involving the nonequilibrium work $W$ and heat exchanged with the hot bath $Q_H$. In a previous work [Sinitsyn N A, J. Phys. A: Math. Theor. {\bf 44} (2011) 405001] this fluctuation relation was obtained under the assumption that the heat engine undergoes a stochastic jump process. Here we provide the general quantum derivation, and also extend it to the case of refrigerators, in which case Carnot's statement reads: $\langle Q_C \rangle / |\langle W \rangle| \leq (T_H/T_C-1)^{-1}$.
\end{abstract}

\maketitle

\section{Introduction}
In the last two decades a great attention has been devoted to the understanding of the microscopic underpinnings of the various formulations of the second law of thermodynamics on the basis on exact non-equilibrium identities, collectively known as fluctuation relations \cite{Seifert08EJPB64,Marconi08PREP111,Esposito09RMP81,Campisi11RMP83,Campisi11RMP83b,Jarzynski11ARCMP2,Seifert12RPP75}. 
For example, when applied to a cyclic process occurring in a system in contact and initially in equilibrium with a bath of temperature $T$, the Jarzynski identity $\langle e^{-W/k_B T} \rangle=e^{-\Delta F/k_B T}$, an exact theorem in Hamiltonian (classical or quantum) mechanics \cite{Jarzynski97PRL78}, implies that the average work done on the system is non-negative $\langle W \rangle \geq 0$, namely no energy can be extracted from the bath (Kelvin formulation of the second law)\footnote{$\Delta F$ denotes the difference between the free energy of the (not necessarily reached) state of thermal equilibrium corresponding to the end-point of the protocol and the initial equilibrium state. For a cyclic protocol $\Delta F=0$. $k_B$ denotes Boltzmann's constant.}. Likewise, Clausius formulation, $\int \delta Q/T \leq \Delta S$ follows from a similar identity $\langle e^{-\Delta(E/k_BT)+ \int dQ/T} \rangle=e^{-\Delta (F/k_BT)}$, that applies when a driven system is brough into contact with various thermal baths \cite{Jarzynski99JSM96}.
When studying the heat $Q$ flowing from a hot bath to a cold bath, with no external driving, the  exchange fluctuation relation \cite{Jarzynski04PRL92,Andrieux09NJP11,Campisi10PRL105} implies the identity $\langle e^{-(\beta_C-\beta_H)Q}\rangle =1$ which in turn implies $\langle Q \rangle\geq 0$, saying that heat, on average, cannot flow from cold to hot: yet another equivalent formulation of the second law.

Similarly, Sinitsyn \cite{Sinitsyn11JPA4} derived a nonequilibrium  identity for heat engines,
$\langle e^{-(\beta_C-\beta_H)Q_H + \beta_C W} \rangle =1$, leading to Carnot's formulation: no heat engine can overperform the Carnot engine, $\langle W \rangle /\langle Q_H \rangle \leq 1- T_C/T_H$, see also Ref. \cite{Lahiri12JPA45}.
These derivations are based on the assumption that the engine dynamics is described by a stochastic jump process obeying detailed balance. Sinitsyn however suggested \cite{Sinitsyn11JPA4} that a more general Hamiltonian derivation should be possible as well. Here we follow that suggestion and provide a proof of the fluctuation relation for heat engines under the assumption that device and baths obey the laws of quantum mechanics.
We further extend that result to the case of reverse operation, namely for quantum refrigerators.

A number of proposals have been recently made of quantum engines and refrigerators, see, e.g. \cite{Abah12PRL109,Venturelli13PRL110,Correa14SciRep4}.
The fluctuation relations presented here constitute a basic theoretical tool for the study of those and similar devices.

\section{Fluctuation relation for quantum heat engines} 
%%%%%%%%%%%%%%%%%%%%%%%%% FIGURE %%%%%%%%%%%
\begin{figure}[]
		\begin{center}
		\includegraphics[width=.3\linewidth]{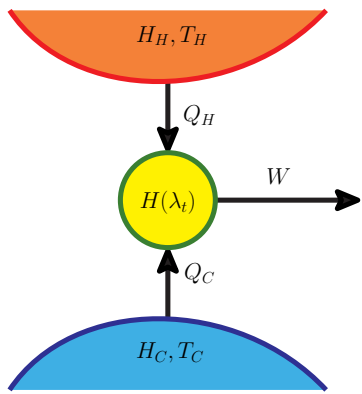}
		\caption{(Color online) Schematics of a heat engine/refrigerator. A device (yellow circle) exchanges the heats, $Q_H, Q_C$, with two thermal reservoirs of temperature $T_H, T_C$, respectively, and outputs the energy $W$. The arrow directions reflect the sign convention adopted in this work. Heat is positive when flows out of a reservoir, work is positive when done on the external body. When the device operates as a heat engine it is $\langle Q_H \rangle \geq 0, \langle Q_C \rangle \leq 0, \langle W \rangle \geq 0$. When the device operates as a refrigerator it is $\langle Q_H \rangle \leq 0, \langle Q_C \rangle \geq 0, \langle W \rangle \leq 0$.}
		\label{fig:Fig1}
		\end{center}
\end{figure}
%%%%%%%%%%%%%%%%%%%%%%%%%%%%%%%%%%%%%%%%%%
Consider a heat engine, Fig. \ref{fig:Fig1}. A device  undergoes a cyclic process during which it exchanges the heats $Q_C, Q_H$ with two thermal reservoires at temperatures $T_C\leq T_H$, respectively, and outputs energy in the form of work $W$. 
We treat the reservoires and device as a compound quantum mechanical system.
Its Hamilton operator reads \cite{Jarzynski99JSM96}:
\begin{equation}
\mathcal H(t) = H(\lambda_t)+ H_H +  H_C+ c_tV_C  + h_t V_H 
\label{eq:Htot}
\end{equation}
Here $H(\lambda_t)$ is the Hamiltonian of the device. It depends on time through the parameter $\lambda$ accounting for the coupling with the external work source/sink. The device can be for example  a gas in a cylindrical container of volume $\lambda$. The device undergoes a cycle of duration $\tau$, i.e. $\lambda_0= \lambda_{\tau}$.
The two thermal reservoires have Hamiltonians $H_{H}, H_C$, while $V_C, V_H$ denote their coupling to the device. 
The time-dependent real coefficients $0\leq c_t, h_t\leq 1$ denote the degree to which each coupling is switched on.
Following \cite{Sinitsyn11JPA4} we consider the case when $c_t=1$ and $h_t=\Theta(t)$ with $\Theta$ the Heaviside step function.
Accordingly the device is prepared at $t<0$ in a state of thermal equilibrium with the cold bath.
Assuming the coupling $V_C$ is weak,  the initial state is given, up to second order corrections in the coupling, by \cite{Talkner09JSM09}:
\begin{equation}
\rho_0= \frac{e^{-\beta_C H(\lambda_0)}}{Z_0(\beta_C)} \otimes \frac{e^{-\beta_C H_C}}{Z_C} \otimes
\frac{e^{-\beta_H H_H}}{Z_H}
\label{eq:initial}
\end{equation}
where $\beta_{C(H)}=(k_B T_{C(H)})^{-1}$, $k_B$ is Boltzmann constant, $Z_0(\beta_C)=\Tr_D e^{-\beta_C H(\lambda_0)}, Z_C = \Tr_C e^{-\beta_C H_C}, Z_H = \Tr_H e^{-\beta_H H_H}$, with $\Tr_{D,H,C}$ denoting the trace over the device, hot bath, and cold bath respective Hilbert spaces. 	According to the two-measurement scheme \cite{Esposito09RMP81,Campisi11RMP83,Kurchan00arXiv,Tasaki00arXiv,DeRoeck04PRE69,Talkner07JPA40},
at time $t=0$ projective measurements are performed of the energy of the device $H(\lambda_0)$, the energy of the cold bath $H_C$, and  the energy of the hot bath $H_H$. Let $E_\nu^0, E_n^C, E^H_N$ denote the respective outcomes. Note that $H(\lambda_0),H_C $ and $H_H$ commute with each other, hence they can be simultaneously measured. The device+baths then evolves up to time $\tau$ according to the unitary evolution generated by the total Hamiltonian (\ref{eq:Htot}). At the final time, $t=\tau$, joint projective energy measurements of $H(\lambda_\tau), H_C$ and $H_H$ are again performed with outcomes $E_\mu^\tau, E_m^C, E^H_M$. Under the assumption that the total Hamiltoinan $\mathcal H(t)$ is time-reversal invariant at each time $t$, the joint probability 
$p(\Delta E, Q_H, Q_C)$ of observing the energy changes $\Delta E = E_\mu^\tau-E_\nu^0$, $Q_C= E_n^C-E_m^C $, $Q_H= E_N^H-E_M^H $ obeys the fluctuation relation \cite{Andrieux09NJP11,Campisi10PRL105}:
\begin{equation}
\frac{p(\Delta E, Q_H, Q_C)}{\widetilde p(-\Delta E, -Q_H, -Q_C)}= e^{\beta_C \Delta E -\beta_C Q_C-\beta_H Q_H}\, .
\label{eq:FT}
\end{equation}
where $\widetilde p(-\Delta E, -Q_H, -Q_C)$ is the probability of measuring
$-\Delta E, -Q_H, -Q_C$ in the reverse operation of the machine $\widetilde{\lambda}_t= \lambda_{\tau-t}$ starting in the initial state $\rho_\tau=e^{-\beta H(\lambda_\tau)}/\Tr e^{-\beta H(\lambda_\tau)}= \rho_0$ (recall that we consider a cyclic driving, $\lambda_0=\lambda_\tau$). The proof of Eq. (\ref{eq:FT}) crucially depends on the unitarity and time reversal symmetry of the system+baths compound.

Because of energy conservation the work output $W$ equals the negative total energy change in the device+baths, hence, neglecting the weak coupling energies:
\begin{equation}
W = Q_C + Q_H - \Delta E \, .
\label{eq:NRGconservation}
\end{equation}
Thus  we arrive at the fluctuation relation for quantum heat engines:
\begin{equation}
\frac{P(\Delta E, Q_H, W)}{\widetilde P(-\Delta E, -Q_H, -W)}= e^{(\beta_C-\beta_H)Q_H - \beta_C W}\, ,
\label{eq:FRQHE}
\end{equation}
where $P(\Delta E, Q_H, W)=p(\Delta E, Q_H, W- Q_H + \Delta E)$ is the joint probability of increasing the system energy by $\Delta E$, exchanging heat $Q_H$ and performing the work $W$, and $\widetilde P(-\Delta E, -Q_H, -W)=\widetilde p(-\Delta E, -Q_H, -W+ Q_H- \Delta E)$ is the joint probability of  increasing the system energy by $-\Delta E$, exchanging heat $-Q_H$ and performing the work $-W$ in the reverse process. Note that the r.h.s. of Eq. (\ref{eq:FRQHE}) does not depend on $\Delta E$. This cancellation was already pointed out in \cite{Sinitsyn11JPA4}, and crucially depends on the fact that the device is prepared in equilibrium with the cold bath. Multiplying Eq. (\ref{eq:FRQHE}) by $\widetilde P(-\Delta E, -Q_H, -W)e^{-(\beta_C-\beta_H)Q_H + \beta_C W}$ and integrating out $\Delta E,Q_H,W$ we obtain the nonequilibrium identity:
\begin{equation}
\langle e^{-(\beta_C-\beta_H)Q_H + \beta_C W} \rangle =1\, .
\label{eq:integratedFRQHE}
\end{equation}
Invoking now Jensen's inequality we obtain:
\begin{equation}
\langle Q_H \rangle (\beta_C-\beta_H) \geq \beta_C \langle W \rangle \, .
\label{eq:ineq}
\end{equation}
Note that $\beta_C-\beta_H\geq 0$. Assuming that the machine operates as a heat engine, it is $\langle Q_H \rangle \geq 0$,
hence we obtain
\begin{equation}
\frac{\langle W \rangle}{ \langle Q_H \rangle} \leq 1-\frac{T_C}{T_H} \qquad $(heat engine)$\, .
\label{eq:2ndLaw}
\end{equation}

\section{Fluctuation relation for quantum refrigerators} 
It is interesting to see what happens when the machine works as a refrigerator. We now consider the case $c_t=\Theta(t), h_t=1$, and prepare the device in a state of thermal equilibrium with the hot bath at time $t<0$:
\begin{equation}
\rho_0= \frac{e^{-\beta_H H(\lambda_0)}}{Z_0(\beta_H)} \otimes \frac{e^{-\beta_C H_C}}{Z_C} \otimes
\frac{e^{-\beta_H H_H}}{Z_H}
\end{equation}
Repeating the same steps as above one arrives at:
\begin{equation}
\frac{P(\Delta E, Q_H, W)}{\widetilde P(-\Delta E, -Q_H, -W)}= e^{(\beta_H-\beta_C)Q_C - \beta_H W}\, .
\label{eq:FRQHE-fridge}
\end{equation}
hence:
\begin{equation}
\langle Q_C \rangle (\beta_H-\beta_C) \geq \beta_H \langle W \rangle\, .
\label{eq:ineq}
\end{equation}
Assuming the engine is working as a refrigerator, it is $\langle W \rangle \leq 0$, hence we arrive at:
\begin{equation}
\frac{\langle Q_C \rangle}{ |\langle W \rangle|} \leq \frac{1}{T_H/T_C-1} \qquad $(refrigerator)$\, .
\label{eq:2ndLawFridge}
\end{equation}

\section{Remarks}
Invoking the classical version of the fluctuation relation
\cite{Jarzynski99JSM96,Jarzynski00JSP98}, which reads exactly as the quantum one, the same results can be obtained within the classical Hamiltonian formalism as well. Note that the interpretation of $Q_H$ $Q_C$ as heat is only possible under the assumption that
the couplings $V_H,V_C$ are weak, at least at the times $t=0,\tau$ of the two measurements \cite{Campisi11RMP83,Talkner09JSM09}.
As already pointed out by Sinitsyn \cite{Sinitsyn11JPA4}, Eq. (\ref{eq:FRQHE}) includes the work fluctuation relation for cyclic processes, and the exchange fluctuation theorem as special cases. The work fluctuation theorem for cyclic processes emerges when $\beta_C=\beta_H=\beta$, namely the system is in contact with a single bath. In this case one obtains $P/\widetilde P = e^{-\beta W}$ \footnote{Note that here work is defined as the energy put into the external bodies, while customarily, in the context of work fluctuation relations, it is defined as the energy put into the system+baths.}. This same relation is also obtained, with $\beta=\beta_C$, if the coupling to the hot bath is switched off, $h_t=0$, or if both couplings are switched off $c_t=h_t=0$. Note that the quantum heat engine fluctuation relation, Eq. (\ref{eq:integratedFRQHE}), remains valid also if only the coupling to the cold bath is switched off. This evidences that what counts the most in obtaining it is the initial preparation of the system in equilibrium with the \emph{cold} bath, and the ability to interact with a hotter bath (vice-versa for a refrigerator).
The exchange fluctuation relation $P/\widetilde P = e^{Q(\beta_C-\beta_H)}$ emerges when there is no device, hence no driving, $W=0$, and no $\Delta E$, namely $Q_H=-Q_C=Q$.

It is important to notice that in a real implementation, where the driving $\lambda_t$ is a periodic function of time, the system typically enters into a non-equilibrium steady state, in which at the beginning of each cycle the system is not necessarily in equilibrium with the cold (or hot) bath. This poses no problem regarding the validity of the here presented results, because all that is necessary for them to hold is that the system is in equilibrium with the cold (or hot) bath at the beginning of the driving, i.e. at the beginning of the very first cycle only.

The physical meaning of the bound in Eq. (\ref{eq:2ndLaw}) is that when 
repeating the cycle a very large number $N$ of times, the total work output $N \langle W \rangle$ over the total heat input $N\langle Q_H \rangle$ cannot overcome Carnot efficiency. Note that the efficiency $\eta=W/Q_H$ of each single operation is a stochastic quantity obeying the statistics
\begin{equation}
\mathcal P(\eta) = \int d\Delta E\, dQ_H\, dW\,  P(\Delta E, Q_H, W) \delta(\eta - W/Q_H)
\end{equation}
where $\delta(x)$ denotes Dirac's delta function.  In general neither the variable $\eta$ nor its average $\langle \eta \rangle = \int d\eta\, \mathcal P(\eta) \eta=\langle W/Q_H \rangle$ (which can well differ from $\langle W\rangle /\langle Q_H \rangle$ due to possible correlations between $Q_H$ and $W$) are bounded by Carnot's efficiency. For recent investigations on the statistics of efficiency see, e.g. \cite{Verley14arXiv,Rana14arXiv}.

\section*{Ackowledgements}
Fruitful discussions with Prof. R. Fazio are gratefully acknowledged. This research was supported by a Marie Curie Intra European Fellowship within the 7th European Community Framework Programme through the project NeQuFlux grant n. 623085 and by the COST action MP1209 ``Thermodynamics in the quantum regime''. 

\section*{References}

\providecommand{\newblock}{}

\end{document}